\documentclass[12pt]{article}
\usepackage{epsfig}
\usepackage{amsfonts}
\usepackage{amsmath}
\usepackage{amssymb}
\usepackage{color} 

\newcommand{\beq}{\begin{equation}}
\newcommand{\eeq}{\end{equation}} 
\def\nuc#1#2{\relax\ifmmode{}^{#1}{\protect\text{#2}}\else${}^{#1}$#2\fi}
\def\itnuc#1#2{\setbox\@tempboxa=\hbox{\scriptsize\it #1}
\def\@tempa{{}^{\box\@tempboxa}\!\protect\text{\it #2}}\relax
\ifmmode \@tempa \else $\@tempa$\fi}

\newcommand{\slashpi}{\slash \hspace{-0.5em}\pi}
\newcommand{\minislashpi}{\slash \hspace{-0.3em}\pi}

\newcommand{\simge}{\hspace*{0.2em}\raisebox{0.5ex}{$>$}
     \hspace{-0.8em}\raisebox{-0.3em}{$\sim$}\hspace*{0.2em}}
\newcommand{\simle}{\hspace*{0.2em}\raisebox{0.5ex}{$<$}
     \hspace{-0.8em}\raisebox{-0.3em}{$\sim$}\hspace*{0.2em}}

\newcommand{\MQCD}{M_{\mathrm{QCD}}}

\newcommand{\dslash}[1]{#1 \llap{/\kern-0.5pt}}
\newcommand{\Dslash}[1]{#1 \llap{/\kern+1.2pt}}
\newcommand{\DDslash}[1]{#1 \llap{/\kern+2.3pt}}
\newcommand{\dslashh}[1]{#1 \llap{/\kern+1pt}}

\newcommand{\boldtau}{\mbox{\boldmath $\tau$}}
\newcommand{\boldpi}{\mbox{\boldmath $\pi$}}

\def\bdm{\begin{displaymath}}
\def\edm{\end{displaymath}}

\begin{document}

\begin{titlepage}

\vspace*{1.5cm}

\begin{center}
{\Large\bf Naturalness in} 
\\
\vspace{0.3cm}
{\Large\bf Nuclear Effective Field Theories}
\\

\vspace{2.0cm}

{\large \bf U. van Kolck}

\vspace{0.5cm}
{
{\it Institut de Physique Nucl\'eaire, CNRS/IN2P3,
\\
Universit\'e Paris-Sud, Universit\'e Paris-Saclay,
\\
91406 Orsay, France}
\\
and
\\
{\it Department of Physics, University of Arizona,
\\
Tucson, AZ 85721, USA}
}

\vspace{1cm}

\today

\end{center}

\vspace{1.5cm}

\begin{abstract}
Nuclear effective field theories (EFTs) have been developed over the
last quarter-century with considerable impact on the description
of light and even medium-mass nuclei.
At the core of any EFT is a systematic expansion
of observables, which is usually obtained from a rule based
on an assumption of naturalness. 
I discuss naturalness in the context of the relatively weak binding
of nuclei, where discrete scale invariance plays a role in the
emergence of complexity.
\end{abstract}

\end{titlepage}

\section{Introduction}
\label{intro}

In particle physics and cosmology, 
one most frequently hears about the concept of naturalness in connection
to problems with the Standard Model (SM) \cite{Dine:2015xga}: 
strong CP, gauge hierarchy, cosmological constant.
It is easy to forget that naturalness is a cornerstone for the paradigm 
to understand the successes of quantum field theory across mass scales,
effective field theories (EFTs) \cite{Weinberg:1978kz}.
The SM is but one example of an EFT, one in which little information is known 
about subleading interactions.
My goal here is to discuss some aspects of naturalness in nuclear EFTs,
where the weak binding of nuclei is also usually seen as a naturalness 
problem, but subleading interactions are important and 
assuming some version of naturalness is crucial for predictions.
 
Nature is organized as a tower of EFTs \footnote{Just so the true
believers burn me at the stake for the right reason: I ain't saying the tower
never ends, only that we'll not know for sure.}, each with 
at least two mass scales: the scale $M_{\rm lo}$
we probe with reactions where the typical external momentum $Q\sim M_{\rm lo}$,
and the breakdown scale 
\footnote{In the particle physics literature, where dimensional regularization
is almost exclusively used, the physical breakdown scale is normally referred to
as the ``cutoff'' of the theory. Unfortunately dimensional regularization
is not well adapted to nonperturbative problems where loops do not factorize,
as in nuclear physics (except for a very specific situation mentioned
below).
Here I reserve ``cutoff'' to
the arbitrary momentum (or coordinate) cutoff introduced by the regularization
procedure.}
$M_{\rm hi}$ where a reorganization of the theory
(new degrees of freedom, reordering of interactions) is needed.
In an EFT, all interactions among the relevant degrees of freedom
are included which are allowed by symmetries,
since even if an allowed interaction
were magically absent at one scale it would still be present at other scales.
The only way to make predictions is to first make an assumption about
the magnitudes of masses and interaction strengths.
Naturalness, in the form of an expectation about the effects of an EFT 
parameter on observables, offers a rule to infer the
hierarchy of interactions (``power counting'').
One usually speaks of fine tuning when naturalness expectations
are not fulfilled and yet no new symmetry is identified.
Naturalness considerations are not idle, as when naturalness
fails for one observable we need to determine the extent of power-counting
revision.

The scarcity of data challenging the SM is probably
the engine that drives the emphasis on naturalness problems.
In contrast, data abound in nuclear physics, which has proven to be fertile 
ground for EFTs (see Ref. \cite{Hammer:2019poc} for a 
recent comprehensive review).
From early on, nuclear theory has been a constant struggle to
explain the many regularities seen among nuclear properties
while facing severe renormalization difficulties.
(For a short history, see Ref. \cite{Machleidt:2017vls}.)
There has always been a feeling, although it is hard to trace
its origins, that ``nuclear physics is fine tuned''.
EFT naturalness provides a framework to address this issue.

It is in fact hard to go anywhere with nuclear EFTs 
without guidance from naturalness. No nuclear EFT can go beyond 
the scale associated with nonperturbative effects
in Quantum Chromodynamics (QCD),
$\MQCD \sim 1$ GeV, which sets the scales
for hadronic masses such as the nucleon mass $m_N\simeq 940$ MeV.
Ground states in heavy nuclei have an approximately
constant binding energy per nucleon $B_A/A$,
which we can associate with a binding momentum
\footnote{This estimate  
gives the correct position of the $T$-matrix pole for $A=2$
and ensures that all nucleons contribute equally to the binding energy
when $A\gg 2$.}
$\kappa_A \sim \sqrt{2m_NB_A/A}$. 
For the alpha particle, frequently considered as a light 
representative of typical nuclei, $\kappa_4 \simeq 100$ MeV.
While these specific estimates for $\MQCD$ and $\kappa_A$ can easily 
be wrong by factors of ${\cal O}(1)$,
it is well recognized that
the bane of nuclear physics is a limited separation of scales.
In contrast with atoms where the Coulomb interaction
is clearly dominant and effects from other electromagnetic forces
are very small, in nuclear physics even the identification of
leading interactions is challenging.

Much nuclear EFT work has been carried out as
an uncritical application of ``naive dimensional analysis''
\cite{Manohar:1983md,Georgi:1986kr,Weinberg:1989dx,Georgi:1992dw}.
As I discuss in Sec. \ref{flavors}, this is a rule
based on the sensitivity of loop diagrams to high-momentum physics,
which applies to a large class of EFTs where all interactions
are perturbative. 
Naive dimensional analysis frames, for example, the SM naturalness debate.
But purely perturbative nuclear physics is not, else there would
be no nuclei. The implications of renormalization in a nonperturbative
context are perhaps the most distinctive feature of nuclear EFTs
--- for a review, see Ref. \cite{vanKolck:2020llt}.

A characteristic feature of nuclear physics is that nucleons
are heavy (compared to their binding momenta) and stable (or nearly so).
Loops are sensitive to $M_{\rm hi}$ in a different way than in 
most known EFTs \cite{Weinberg:1990rz,Weinberg:1991um,vanKolck:1997ut,
Kaplan:1998tg,Kaplan:1998we,vanKolck:1998bw}.
The appropriate modification of naive dimensional analysis
is introduced in Sec. \ref{nonpert},
and corroborated by specific examples.
The idea that nuclear physics is fine tuned might be traced back to
the fact that,
even with such modification, light nuclei are unnaturally shallow.

In Sect. \ref{symmetry} I argue that this unnaturalness
can be explained away, at least for light nuclei, by a symmetry 
--- discrete scale invariance --- 
whose importance has not been fully appreciated until recently
\cite{Konig:2015aka,Konig:2016utl,Konig:2016iny,Kolck:2017zzf,Konig:2019xxk}.
Discrete scale invariance emerges within
the nuclear EFT designed to deal with light nuclei, Pionless EFT,
in the unitarity limit where two-body binding energies vanish. 
This symmetry allows for growing complexity as $A$ increases,
including some threshold coincidences that one would
have thought arise from delicate fine tuning. 

While it is not obvious that this description can be extended to 
heavier nuclei, if it does one could say nuclear physics is technically 
natural. Some unresolved issues with this conjecture 
are present in Sec. \ref{conc} in guise of a conclusion.

These ideas have been explored separately before and are
of course known to be related. My intention in bringing them
together here is not to contribute to the history of the concept of 
naturalness and to the contemporary preoccupation with the gauge 
hierarchy problem.
Many of the interesting philosophical underpinnings
of naturalness have been examined in, for example,
Refs. \cite{Nelson:1985,Giudice:2008bi,Grinbaum:2009sk,Wells:2013tta,Williams:2015gxa,Hossenfelder:2018ikr,Bain:2019,Wells:2018sus,Williams:2018ppw,Borrelli:2019uab}.
My goal is instead to highlight the connection between the 
general ideas that permeate these discussions
and a specific situation where naturalness (or lack thereof)
has very observable consequences.

\section{Enunciating naturalness}
\label{flavors}

The idea that ``fundamental'' parameters should not differ significantly
in magnitude has been around for a long time, being
expressed in a particularly
clear form by Dirac \cite{Dirac:1937ti}.
In an EFT, fundamental parameters are replaced  in an effective Lagrangian
by an infinite number of interaction strengths, 
known as low-energy constants (LECs) or Wilson coefficients. 
These parameters are in general dependent on the regulator
one introduces to make sense of quantum corrections; 
they are ``bare'', unobservable parameters. 
In observables, they always appear together with the high-momentum components
of quantum corrections,  which also depend on the regulator,
in what are frequently referred to as ``renormalized'' parameters.
EFT is all about observables, and when effective-field theorists
talk about naturalness, they are thinking about the size of the
renormalized LEC 
\footnote{When nonperturbative physics is involved, the calculation
of quantum corrections is often limited to numerics and
one cannot write a simple analytical formula for the renormalized LEC.
In nuclear physics, this limitation causes an inordinate amount of confusion.}.
More precisely, naturalness concerns the
effect the finite part of a LEC has on observables.

The simplest version of naturalness is probably plain
dimensional analysis, where any LEC associated with 
an operator of mass dimension $D$ appearing in the Lagrangian
is assumed to be 
\begin{equation}
c_D= {\cal O}(M_{\rm hi}^{4-D}).
\label{plainNDA}
\end{equation}
While frequently this is enough for rough estimates,
there are many instances where it is not.
We would like to account for possible small coupling constants.
And, even when the theory is strongly coupled, extra factors of
$4\pi$ can be essential in providing a low-energy scale $M_{\rm lo}$.

As an example relevant to nuclear physics, let us consider 
two-flavor Chiral EFT ($\chi$EFT)  \cite{Weinberg:1978kz,Gasser:1983yg}
--- the EFT of QCD where $M_{\rm hi}={\cal O}(\MQCD)$. 
It includes an isotriplet of pions $\boldpi$
with a root-mean-square mass $m_\pi\simeq 137.3$ MeV 
and charge-neutral squared-mass splitting
$\delta m_\pi^2 \equiv m_{\pi^\pm}^2-m_{\pi^0}^2\simeq (35.5 \, \mathrm{MeV})^2$,
and, because they are (at least, nearly) stable, 
an isodoublet of nucleons $N$ with average mass 
$m_N=(m_p+m_n)/2\simeq 938.9$ MeV
and neutron-proton mass splitting $\delta m_N\equiv m_n-m_p\simeq 1.3$ MeV.
With $\vec\sigma$ ($\boldtau$) denoting the Pauli matrices
in spin (isospin) space, the Lagrangian can be written 
\begin{eqnarray}
{\cal L}_{\chi{\rm EFT}} &=& 
\frac{1}{2}\left[(\partial_0 \boldpi)^2 -(\vec\nabla \boldpi)^2 
- m_\pi^2\boldpi^2 
- \delta m_\pi^2\left(\boldpi^2-\pi_3^2\right)\right]
\nonumber\\ 
&&
+ N^\dagger \left(i \partial_0 + \frac{\vec \nabla^2}{2m_N} 
+ \frac{\delta m_N}{2}\tau_3\right) N
+ \frac{g_A}{2f_\pi} N^\dagger\boldtau \vec\sigma N \cdot \vec\nabla\boldpi
+ \ldots ,
 \label{chiLag}
\end{eqnarray}
where $f_\pi\simeq 92$ MeV is the pion decay constant and
$g_A\simeq 1.27$ is the nucleon axial coupling.
The ``$\ldots$'' represent terms with more fields and derivatives,
as well as further isospin-breaking and weaker interactions.
Plain dimensional analysis gives 
\begin{equation}
m_\pi^2\sim |\delta m_\pi^2| ={\cal O}(\MQCD^2), 
\quad
m_N \sim |\delta m_N| ={\cal O}(\MQCD),
\quad
|g_A|={\cal O}(f_\pi/\MQCD),
\label{plainDA}
\end{equation}
which would suggest that at least
$m_\pi$, $\delta m_\pi^2$, $\delta m_N$, and $g_A^{-1}$ are observed
to be unnaturally small,
an unlikely scenario.
Clearly we should be able to do better. 
The existence of a gap between $f_\pi$ and the masses of most hadrons set by
$\MQCD$ means that, once a better guess is employed for what is natural,
we should see $\chi$EFT as an EFT where $M_{\rm lo}={\cal O}(f_\pi)$. 
In addition, other low-energy scales are present in the form
of $m_\pi$, $\delta m_\pi^2$, and $\delta m_N$. 

More generally, the issue is that LECs can also depend on  
the ratio $M_{\rm lo}/M_{\rm hi}$, since both
scales emerge from the same theory at momenta above $M_{\rm hi}$.
Two essential ideas are:
\begin{itemize}
\item As stressed by Veltman \cite{Veltman:1980mj},
we must consider renormalization explicitly and use loop
corrections to estimate the importance of a LEC.
The actual values of loop integrals
depend on the regulator parameter $\Lambda$
that we introduce to cut large momenta off.
Once $\Lambda \simge M_{\rm hi}$ there is no hope that we can account
for this physics in detail. But this is also physics that
can be mocked up by short-range interactions, and it is only the
combination of bare LECs and high momentum in loops that enters
observables. Renormalization is the demand that the $\Lambda$ dependence
in the bare LECs cancels out that from the loops, no matter how large
$\Lambda$ is. But the dependence on $\Lambda$ from the loops
indicates the sensitivity of the observable to high-momentum
physics that enters at $M_{\rm hi}$. Thus, barring cancelations in
this physics, the replacement $\Lambda \to M_{\rm hi}$ gives
an indication of the natural size of the observable, controlled
by the finite part of the corresponding LEC. 
(For a clear discussion in the context of the cosmological
constant, see Ref. \cite{Burgess:2013ara}.)
This requirement ends up bringing to Eq. \eqref{plainNDA}
additional factors of $4\pi$, arising from angular integration.

\item 't Hooft \cite{tHooft:1979rat}
pointed out that small LECs are unlikely to arise
in the EFT from cancelations
among parameters in the underlying theory, and are expected instead
to come from symmetries.
An exact symmetry will guarantee that an infinite number of 
otherwise possible LECs
vanish, as do loop corrections to their tree-level values
(when a symmetry-preserving regulator is used). 
An approximate symmetry will allow both bare LECs and
non-vanishing loop corrections,
but they will be suppressed by powers of the small symmetry-breaking 
parameters: the LECs are ``protected'' from
receiving large corrections.
This idea is frequently referred to as ``technical'' naturalness.
Equation \eqref{plainNDA} should be modified to account for
parameters that break symmetries gently.
\end{itemize}

In perturbation theory, these two ideas are incorporated in naive 
\footnote{``Naive'' perhaps due to the modesty of the authors of
Ref. \cite{Manohar:1983md}.}
dimensional analysis (NDA) 
\cite{Manohar:1983md,Georgi:1986kr,Weinberg:1989dx,Georgi:1992dw}.
Perhaps the simplest version is that of Weinberg's \cite{Weinberg:1989dx},
where the LEC of an operator involving $N$ fields is natural if 
\begin{equation}
c_{DN} = {\cal O}\left((4\pi)^{N-2} \, M_{\rm hi}^{4-D} \, c_{DN}^{\text{red}}\right)
\label{WNDA}
\end{equation}
in terms of the dimensionless ``reduced'' LEC $c_{DN}^{\text{red}}$.
When the underlying theory
has no small parameters, $c_{DN}^{\text{red}}={\cal O}(1)$. 
When it does, the size of the reduced LEC is
set by the minimum powers of the (small) reduced parameters
associated with the underlying operators
needed to generate the corresponding operator in the EFT.
We can now schematically 
write a general term in an effective Lagrangian containing 
$b$ (relativistic) bosonic fields $\phi$, 
$f$ fermionic fields $\psi$,
and $d$ derivatives as 
\cite{Manohar:1983md,Georgi:1986kr,Georgi:1992dw}
\begin{equation}
{\cal L}_{\rm EFT} \sim
(4\pi)^{-2} M_{\rm hi}^4 \; c_{DN}^{\text{red}}
\left(\frac{\partial}{M_{\rm hi}}\right)^d
\left(\frac{4\pi \phi}{M_{\rm hi}}\right)^b 
\left(\frac{(4\pi)^2\bar\psi \psi}{M_{\rm hi}^3}\right)^{f/2},
\label{GNDA}
\end{equation}
using $N=b+f$ and $D=b+3f/2+d$.
Compared to plain dimensional analysis, there are extra factors
of $4\pi$ and, per technical naturalness, $c_{DN}^{\text{red}}$.

The improvement over plain dimensional analysis is clear in $\chi$EFT. 
The underlying theory is QCD with additional electromagnetic
interactions given by QED and weaker interactions given by Fermi theory.
For two flavors --- up and down quarks of masses $m_u$ and $m_d$ ---
the small parameters are the reduced average quark mass
$\hat{m}^{\text{red}}=\hat{m}/\MQCD\equiv (m_u + m_d)/2\MQCD\sim 0.003$,
the reduced down-up quark-mass difference 
$(\epsilon \hat{m})^{\text{red}}= (m_d - m_u)/2\MQCD \sim \hat{m}^{\text{red}}/3$,
the reduced electromagnetic coupling 
$e^{\text{red}}= \sqrt{\alpha_e/4 \pi}\sim 0.02$,
and reduced parameters associated with weaker interactions.
The smallness of the quark masses is technically natural because
when they are zero the theory has a chiral symmetry of independent
rotations of left- and right-handed quarks.
In two-flavor QCD this is $SU(2)_L\times SU(2)_R \sim SO(4)$. 
Since only the vector $SU(2)_V \sim SO(3)$ subgroup is realized 
in the spectrum, 
chiral symmetry must be spontaneously broken, leading to the appearance 
of three pseudo-Goldstone bosons that parametrize the coset space
$SO(4)/SO(3)\sim S^3$, a 3-sphere.
Chiral symmetry is explicitly 
broken not only by the quark masses but also
by electromagnetic and weaker interactions.
Identifying the pseudo-Goldstone bosons as the three pions,
we can write the interactions in the ``\ldots'' of
Eq. \eqref{chiLag} so that $SU(2)_L\times SU(2)_R$ is broken just
as in the underlying theory.
For example, for the chirally symmetric pion kinetic term,
\begin{equation}
\left[(\partial_0 \boldpi)^2 -(\vec\nabla \boldpi)^2\right]
\to 
\left[(\partial_0 \boldpi)^2 -(\vec\nabla \boldpi)^2\right]
\left(1-\frac{\boldpi^2}{2f^2}+\ldots\right),
\label{chiralpartners}
\end{equation}
where $f$ is the bare pion decay constant.
Equation \eqref{WNDA} now gives, instead of Eq. \eqref{plainDA},
\begin{eqnarray}
&m_\pi^2={\cal O}(\hat{m}\MQCD),
\quad
|\delta m_\pi^2|={\cal O}(\alpha_e\MQCD^2/4\pi),&
\nonumber\\
&m_N ={\cal O}(\MQCD),
\quad
|\delta m_N| ={\cal O}(\epsilon \hat m,\alpha_e\MQCD/4\pi),&
\nonumber\\
&f_\pi={\cal O}(\MQCD/4\pi),
\quad
|g_A|={\cal O}(1).
&
\label{chEFTNDA}
\end{eqnarray}
These estimates work within a factor of $\sim 2$, yielding
$m_\pi^2\sim (70 \; \mathrm{MeV})^2$, 
$\delta m_\pi^2\sim (30 \; \mathrm{MeV})^2$,
$m_N \sim 1$ GeV,
$\delta m_N \sim 1$ MeV,
and 
$f_\pi\sim 80$ MeV.
The naturalness problems of plain dimensional analysis
have been explained away by factors of $4\pi$ (in the case of $g_A$)
and, additionally, by approximate
chiral symmetry (for $m_\pi^2$, $\delta m_\pi^2$, and $\delta m_N$).

This is a case where we could infer the existence
of an approximate symmetry from a pattern of violation of naturalness,
but it is not the only one.
In another beautiful example,
the magnitudes of lepton masses 
yield clues about the SM gauge structure \cite{Wells:2013tta}.
Unfortunately,
similar paths emerging from the smallness of the QCD vacuum angle,
the Higgs mass parameter, and the cosmological constant
are yet to be confirmed experimentally.

What is the basis for NDA?
The various elements that enter Feynman diagrams constructed
from the Lagrangian \eqref{GNDA} are
\begin{eqnarray}
 \text{loop integral} &\sim& (4\pi)^{-2}\, Q^4,
\label{loopintegral}\\
 \text{fermion, boson propagator} &\sim& Q^{-1}, Q^{-2},
\label{props}\\
 \text{vertex} &\sim& (4\pi)^{N-2} \, M_{\rm hi}^{4-D} \, c_{DN}^{\text{red}} \, Q^d,
\label{vertices}
\end{eqnarray}
in terms of the typical external momentum $Q\sim M_{\rm lo}$.
In a diagram with $L$ loops, 
these rules and standard graph identities 
can be used to show that the contribution to $c_{DN}^{\text{red}}$
is ${\cal O}(\Lambda^{2L}/M_{\rm hi}^{2L})$ times the product
of reduced couplings from the various vertices. 
If all the reduced couplings are ${\cal O}(1)$,
then $\Lambda \to M_{\rm hi}$ implies $c_{DN}^{\text{red}}={\cal O}(1)$.
NDA is self-consistent in perturbation theory
\cite{Manohar:1983md}.

As a specific example, consider pion-pion scattering \cite{Manohar:1983md}.
Schematically, the tree amplitude is
\begin{equation}
T_{\pi\pi}^{(L=0)}(Q;\Lambda)\sim \frac{Q^2}{f^2(\Lambda)}
\left[1+ c_{44}^{\rm red}(\Lambda)\frac{Q^2}{M_{\rm hi}^2}+\ldots\right],
\label{Tpipinoloop}
\end{equation}
where the first term comes from the $b=4$ interactions arising
from chiral symmetry (for $d=2$, see Eq. \eqref{chiralpartners}),
while the second term is from $b=4$ interactions
in the ``$\ldots$'' of Eq. \eqref{chiLag}.
Now, a generic one-loop diagram where the
two vertices stem from the former interaction contains
\begin{equation}
T_{\pi\pi}^{(L=1)}(Q;\Lambda)\sim \frac{Q^2}{(4\pi)^2f^4(\Lambda)}
\left[\Lambda^2 + Q^2\ln(\Lambda/m_\pi)\right].
\label{Tpipioneloop}
\end{equation}
The most severe cutoff dependence is removed by the first term
in Eq. \eqref{Tpipinoloop},
\begin{equation}
f_\pi^2\sim f^2(\Lambda) + \frac{\Lambda^2}{(4\pi)^2}+\ldots
\end{equation}
The replacement $\Lambda \to M_{\rm QCD}$ leads to the estimate for 
$f_\pi$ in Eq. \eqref{chEFTNDA}. The
$\ln \Lambda$ can then be absorbed by the second term 
in Eq. \eqref{Tpipinoloop}. 

The situation is analogous for the Higgs mass,
\begin{equation}
m_H^2\sim m^2(\Lambda) + c_2 \, \Lambda^2/(4\pi)^2
+ c_0 \, m^2(\Lambda) \ln(\Lambda/m(\Lambda)) +\ldots
\end{equation}
where $c_{0,2}$ are combinations of dimensionless coupling constants,
which leads to the expectation $m_H={\cal O}(M_{\rm BSM}/4\pi)$ 
in terms of the SM breakdown scale $M_{\rm BSM}$. 
Contrast this with the average pion mass, where 
only the $\ln \Lambda$ appears thanks to chiral symmetry 
\cite{Charap:1970xj,Honerkamp:1996va,Gerstein:1971fm}.
Electromagnetic interactions break isospin and $\delta m_\pi^2$
does receive a $\Lambda^2$ contribution proportional to $\alpha_e/4\pi$.
Again $\Lambda \to M_{\rm QCD}$ leads to the successful estimate
\eqref{chEFTNDA}. This example is in fact often invoked 
(for example, Refs. \cite{Giudice:2008bi,Grinbaum:2009sk}) in
connection to the Higgs mass.

We have seen how NDA provides a benchmark against which to 
measure unnaturalness. From a reductionist perspective,
the hope is that deviations will give clues about the
underlying theory. From the point of view of emergence,
in contrast, NDA offers a basis for the organization of interactions
in the EFT itself, or power counting. 
In order to make predictions we need an amplitude at 
$Q\sim M_{\rm lo}$ to be expressed in rough form as
\begin{equation}
T(Q\sim M_{\rm lo})\sim {\cal N}\sum_{\nu=0}^\infty 
\left(\frac{Q}{M_{\rm hi}}\right)^\nu F_{\nu}(Q/M_{\rm lo}),
\label{genamp}
\end{equation}
where $\nu$ is a counting index,
${\cal N}$ is a process-specific normalization,
and
$F_{\nu}(Q/M_{\rm lo})$ is a combination of LECs and non-analytic functions
stemming from loop integrals. 
The most important terms --- the leading order (LO) --- have $\nu=0$, 
first corrections --- next-to-leading order (NLO) ---
have $\nu=1$, and so on.

In the absence of any small parameters, {\it i.e.}, when 
all $c_{DN}^{\text{red}}={\cal O}(1)$, 
the low-energy scale is $M_{\rm lo}={\cal O}(M_{\rm hi}/4\pi)$.
A diagram with $B$ ($F$) external boson (fermion) lines,
$L$ loops, and
$V_i$ vertices involving $d_i$ derivatives
and $f_i$ fermion fields
will then contribute at \cite{Weinberg:1978kz}
\begin{equation}
\nu = 2L+\sum_i V_i \left( d_i + f_i/2 -2\right), 
\label{WPC}
\end{equation}
while
\begin{equation}
{\cal N}\sim\left(\frac{M_{\rm hi}}{4\pi}\right)^{4-3F/2-B}.
\label{norm}
\end{equation}
The factor of $2L$ in Eq. \eqref{WPC} expresses
the perturbative character of the amplitude.
For example, not only are the higher-derivative terms
suppressed in Eq. \eqref{Tpipinoloop}, but also the non-analytic part of the
loop \eqref{Tpipioneloop} can be treated perturbatively at N$^2$LO.

When an approximate symmetry leads to some $c_{DN}^{\text{red}}\ll 1$, 
they should be compared with 
$M_{\rm lo}/M_{\rm hi}={\cal O}((4\pi)^{-1})$ so as to improve 
the power counting \eqref{WPC}.
Again $\chi$EFT provides an explicit illustration.
Since $\hat{m}^{\text{red}}={\cal O}(m_\pi^2/\MQCD^2)$,
the effect of explicit chiral-symmetry breaking from quark masses
is comparable to that of a derivative (Eq. \eqref{GNDA})
for $Q\sim m_\pi$. 
In the physical world $m_\pi$ is not very different from $f_\pi$, so
we can for simplicity lump them together into $M_{\rm lo}$.
The net effect on the power counting 
is that we can keep Eq. \eqref{WPC} as long 
as $d_i$ now counts powers of $m_\pi$ as well \cite{Weinberg:1978kz}. 
Similar steps can be taken for $(\epsilon\hat m)^{\text{red}}$ and
$e^{\text{red}}$.
 
But what if one or more of the LECs are unnatural after the existing
symmetries are accounted for? 
The obvious guess is that an unknown symmetry is at play, in which case
several LECs are likely affected. 
From the EFT perspective, the task is to include the unnatural
LEC(s) in the power counting and go on with life. 
(If we keep having to repeat this procedure we might
be seeing signs of a previously unknown low-energy degree of freedom,
which again can be incorporated in the EFT, and a new power counting
must be devised. The do-loop starts again.)
We now look at how these ideas play out in nuclear physics.

\section{Nuclear unnaturalness?}
\label{nonpert}

Is nuclear physics unnatural? In this section I discuss some of 
the evidence that suggests the answer might be ``yes''.
I will argue that a small scale enters at the few-nucleon level,
which could then spread through heavier systems.

Before anything, though, we need to realize that NDA in the above
form need not apply.
As it stands, NDA has provided a successful basis for $\chi$EFT in processes
with $Q\sim m_\pi$ where at most one nucleon
is present ($A\le 1$) \cite{Bernard:2007zu}.
It is tempting to use it as a basis to study reactions and
structure of systems with $A\ge 2$,
but this attempt fails from the get-go.
The reason lies on the existence in processes involving only heavy stable 
particles of ``reducible'' diagrams 
containing intermediate states devoid of light particles:
\begin{itemize}
\item
The argument for NDA discussed above attributes $Q^{-1}$
to a fermion propagator, Eq. \eqref{props}.
This is a good estimate for a relativistic fermion, or for 
a single fermion of mass $m_\psi\simge M_{\rm hi}$ interacting with an
external relativistic boson.
In both cases, the energy of the fermion in an intermediate state
is typically ${\cal O}(Q)$.
For a relativistic fermion, the 3-momentum is comparable. 
In the nonrelativistic case, the recoil is only ${\cal O}(Q^2/2m_\psi)$
and can be treated perturbatively, leaving the fermion as static at LO.
However, if recoil were neglected in {\it reducible} diagrams, 
there would be infrared (IR) divergences.
When recoil is retained, there is an IR enhancement ${\cal O}(m_\psi/Q)$ 
compared to the loops considered in the original NDA
\cite{Weinberg:1990rz,Weinberg:1991um}.
\item
In integrals involving relativistic propagators, the typical factor 
resulting from angular integration is the $(4\pi)^{-2}$ 
in Eq. \eqref{loopintegral}. For integrals where one picks
the pole from a heavy particle propagator, though,
there is usually an extra factor of $4\pi$ coming from 
the contour integration over the magnitude of the 3-momentum
\cite{vanKolck:1997ut,Kaplan:1998tg,Kaplan:1998we,vanKolck:1998bw}. 
\end{itemize}
The net effect is to replace Eqs. \eqref{loopintegral} 
and \eqref{props} with
\begin{eqnarray}
\text{reducible loop integral} &\sim& (4\pi m_\psi)^{-1} Q^5,
\label{red2}\\
\text{fermion propagator} &\sim& m_\psi Q^{-2}.
\label{red1}
\end{eqnarray}
A modified form of NDA results if we use these rules with the same
rationale as before:
we find the regulator cutoff dependence of an arbitrary loop
and use $\Lambda \to M_{\rm hi}$
to estimate the magnitude of the contribution from the related
LECs to an observable.

The consequences for naturalness can be seen most easily in 
Pionless EFT ($\slashpi$EFT)
\cite{Hammer:2019poc}, a simpler nuclear EFT 
than $\chi$EFT, where $M_{\rm hi} \sim m_\pi$.
Pions are integrated out, their effects being entirely encoded through LECs.
If, for simplicity of notation, I leave the various spin-isospin factors
implicit,
the effective Lagrangian can be written as 
\begin{eqnarray}
{\cal L}_{\minislashpi\rm{EFT}} &=& 
N^\dagger \left(i \partial_0 + \frac{\vec \nabla^2}{2m_N} 
\right) N
-\frac{2\pi}{m_N}\left\{C_{0}\left(N^\dagger N \right)^2
+C_{2} \left[\left(N^\dagger N \right)\left(N^\dagger \vec\nabla^2 N \right)
+ \mathrm{H.c.}\right]\right\}
\nonumber\\
&&
- \frac{(4\pi)^2}{6m_N} \,  D_0 \left(N^\dagger\!N\right)^3 
- \frac{(4\pi)^3}{8m_N} \,  E_0  \left(N^\dagger\!N\right)^4 
+ \ldots ,
 \label{pilessLag}
\end{eqnarray}
where $C_{0}$, $C_{2}$, $D_0$, and $E_0$ are LECs.
Again, the ``$\ldots$'' represent terms with more fields and derivatives.
For a pedagogical introduction to $\slashpi$EFT, see 
Ref. \cite{vanKolck:2019vge}.

Life is easier in $\slashpi$EFT because {\it all} loops are reducible.
In a case like this, we can replace Eq. \eqref{WNDA} with
\begin{equation}
c_{DN} = {\cal O}\left((4\pi)^{N/2-1} \, m_\psi^{-1} \, M_{\rm hi}^{5-D} \, 
c_{DN}^{\text{red}}\right),
\label{modNDAc}
\end{equation}
Eq. \eqref{GNDA} with
\begin{equation}
{\cal L}_{\rm EFT} \sim
(4\pi)^{-1} \, m_\psi^{-1}\, M_{\rm hi}^5 \, c_{DN}^{\text{red}}
\left(\frac{\partial}{M_{\rm hi}}\right)^d
\left(\frac{4\pi \bar\psi \psi}{M_{\rm hi}^3}\right)^{f/2},
\label{modNDA}
\end{equation}
and Eq. \eqref{vertices} with
\begin{equation}
\text{vertex} \sim (4\pi)^{N/2-1} \, m_\psi^{-1} \, M_{\rm hi}^{5-D} \, 
c_{DN}^{\text{red}} \, Q^d,
\label{red3}
\end{equation}
where again $c_{DN}^{\text{red}}$ is built from the reduced couplings
in the underlying theory.
Explicitly, in the absence of approximate symmetries,
\begin{equation}
|C_{n}|={\cal O}\left(M_{\rm hi}^{-(1+n)}\right),
\quad
|D_{n}|={\cal O}\left(M_{\rm hi}^{-(4+n)}\right),
\quad
|E_{n}|={\cal O}\left(M_{\rm hi}^{-(7+n)}\right),
\label{modNDANN}
\end{equation}
with the usual understanding that this represents the expected
contribution of the corresponding interactions 
to observables after renormalization.

As an example, let us consider the two-nucleon system 
in a specific isospin channel, 
as we did pion-pion scattering earlier.
The on-shell tree-level amplitude is
\begin{equation}
T_{N\!N}^{(L=0)}(k;\Lambda)=-\frac{4\pi}{m_N} 
\left(C_{0}(\Lambda) - k^2 C_{2}(\Lambda) +\ldots\right),
\label{Tnoloop}
\end{equation}
where $k=\sqrt{m_N E}$ in terms of the center-of-mass energy $E$.
The one-loop diagram 
which involves two successive contact interactions with the LEC $C_{0}$ 
gives
\begin{equation}
T_{N\!N}^{(L=1)}(k;\Lambda) = 
\frac{4\pi}{m_N} C_{0}^{2}(\Lambda)
\left(\theta_1 \Lambda + ik
+\theta_{-1} \frac{k^2}{\Lambda} +\ldots \right),
\label{Toneloop}
\end{equation}
where
$\theta_{1-2n}$, $n=0, 1, \ldots$, are numbers that depend on the 
specific form of the (non-local) regulator \cite{vanKolck:1998bw}. 
With this type of regulator, the two-loop diagram with three $C_{0}$s
factorizes. Including this diagram,
the amplitude becomes
\begin{equation}
T_{N\!N}=\frac{4\pi}{m_N}
a_{2} \left[1- ik a_{2} - k^2 a_{2}^2 \left( 1-\frac{r_{2}}{2a_{2}} \right) 
+\ldots\right]
\label{TNNpert}
\end{equation}
after we define the inverse scattering length and the effective range, 
respectively
\begin{equation}
a_{2}^{-1}=C_{0}^{-1}(\Lambda) + \theta_1 \Lambda +\ldots,
\quad
r_{2}=-2\, C_{0}^{-2}(\Lambda)\, C_{2}(\Lambda) - \theta_{-1} \Lambda^{-1} +\ldots
\label{EREparamsEFT}
\end{equation}
Modified NDA translates into
$|a_{2}^{-1}|\sim |r_{2}^{-1}| ={\cal O}(M_{\rm hi})$.
The suppression of the nonanalytic term in Eq. \eqref{Toneloop}
by one power of $Q/M_{\rm hi}$ is different than
the suppression by $(Q/\MQCD)^2$ one finds when $A\le 1$
(cf. Eq. \eqref{Tpipioneloop}),
but the effective-range correction is of relative ${\cal O}(Q^2/M_{\rm hi}^2)$,
as might have been expected from the two powers of $k$.
One can repeat the argument for more-derivative operators
and higher effective-range expansion parameters \cite{vanKolck:1998bw},
recovering Eq. \eqref{modNDANN} for $C_{n}$.

To confirm this naturalness expectation we can consider as underlying
theory an arbitrary potential of range $R\equiv M_{\rm hi}^{-1}$.
Take, say,
a three-dimensional spherical well \cite{vanKolck:1998bw,Braaten:2004rn}
with dimensionless depth $\alpha$,
\begin{equation}
V(\vec{r}) = - \frac{\alpha^2}{m R^2} \, \theta(R-r).
\label{squarewell}
\end{equation}
Solving the Schr\"odinger equation 
for the $S$ wave and expanding the corresponding 
$T$ matrix in powers of $kR$, one finds the effective range parameters
\begin{equation}
a_2 = R \left( 1- \frac{\tan\alpha}{\alpha}\right),
\quad
r_2= R \left( 1- \frac{R}{\alpha^2a_2}-\frac{R^2}{3a_2^2}\right),
\quad
\ldots
\label{EREparamsquarewell}
\end{equation}
For generic values of $\alpha$, we see that
$|a_2|\sim |r_2| \sim \ldots \sim R$ as given by modified NDA.
A plot of $a_2/R$ and $r_2/R$ as functions of $\alpha$ \cite{Braaten:2004rn}
shows that indeed these are the most common values. 
One way to quantify this 
is to assume equal probability for $\alpha$, which translates into
a probability distribution for $a_2/R$ \cite{Braaten:2004rn},
\begin{equation}
p(a_2/R)= \left(\pi \alpha\right)^{-1} \left[(a_2/R-1)^2+\alpha^{-2}\right]^{-1},
\label{sqwella}
\end{equation}
a plot of which shows a prominent peak at 
the natural value $a_2=R$.

Natural contact interactions with size \eqref{modNDANN}
cannot generate bound states, as they are nonperturbative
only for momenta comparable to the breakdown scale. 
That is a problem for nuclear physics,
considering that empirically the two-nucleon
amplitude has shallow poles in both $S$ channels:
an isospin-singlet (triplet) $I=0$ ($I=1$) bound (virtual) state with imaginary
momentum $i\kappa_{20}$ ($i\kappa_{21}$), where 
$\kappa_{20} \simeq a_{20}^{-1}\simeq 45$ MeV
($\kappa_{21}\simeq a_{21}^{-1}\simeq -8$ MeV).
Both binding momenta, especially $\kappa_{21}$, 
are unnatural compared with the expectation from 
modified NDA, $|a_{2I}^{-1}|={\cal O}(m_\pi)$. 

This unnaturalness could be a reflection of fine tuning in the underlying 
theory.
Again, the simple toy model \eqref{squarewell}
provides an example \cite{vanKolck:1998bw}:
when $\alpha \simeq (2n+1)\pi/2 \equiv \alpha_{{\rm c}n}$ with
$n\ge 0$ an integer, we have 
$|a_2|\simeq R |\alpha_{{\rm c}n}(\alpha - \alpha_{{\rm c}n})|^{-1} \simeq 
|\kappa_2^{-1}|\gg R$ 
while still $|r_2|\sim \ldots \sim R$.
For $\alpha$ just below $\alpha_{{\rm c}n}$, there is a shallow virtual state.
As the attraction increases past $\alpha_{{\rm c}n}$, 
a shallow bound state appears. 
One can think of $a_2$ as the size of the bound state, which
in quantum (in contrast to classical) 
mechanics can exceed the range of the potential
\footnote{Note that this is not the only possible fine tuning in this toy model.
One can also make $-r_0/R$ large by fine tuning $a_2/R$ to be small,
that is, dialing a zero of the amplitude to the threshold region.
The low-energy EFT for this situation is a Pionless EFT with
a different scaling of $C_{n}$ 
\cite{vanKolck:1998bw}
than discussed in the following.}.
This example evokes a real-life instance of fine tuning in atomic physics
--- a Feshbach resonance \cite{RevModPhys.82.1225}.
Some atomic systems consist of two coupled channels
with different spin alignments and thresholds.
Variation of an external magnetic field can bring
a bound state in the closed channel to the open-channel threshold.
The scattering length in the lower, open channel becomes arbitrarily large
as the critical value of the magnetic field is approached.
The knob that controls the magnetic field replaces $\alpha$.

Regardless of the reason for the appearance of shallow $S$-wave poles
in the $T$ matrix, $\slashpi$EFT can describe them if after renormalization
\begin{equation}
|C_{n}|={\cal O}\left(|\kappa_2^{-(1+n/2)}|\, M_{\rm hi}^{-n/2}\right),
\label{fineDANN}
\end{equation}
with $|\kappa_2|\ll M_{\rm hi}$.
For $M_{\rm hi}\simge |k|\simge |\kappa_2|$, all diagrams made out of $C_{0}$ 
are comparable and must be resummed.
With a non-local regulator this can be done analytically,
resulting in the LO amplitude
\begin{equation}
T_{N\!N}^{(0)}(k;\Lambda)=\frac{4\pi}{m} \left(\kappa_2 + ik\right)^{-1},
\label{Tsum}
\end{equation}
where 
\begin{equation}
\kappa_2= C_{0}^{-1}(\Lambda)+ \theta_1\Lambda
\label{C00}
\end{equation}
and I dropped terms that can be made arbitrarily small for arbitrarily
large $\Lambda$. Unnaturalness comes from 
the failure of the replacement $\Lambda \to M_{\rm hi}$ as an estimate for 
$\kappa_2$.

A fully systematic description emerges of the low-energy
two-nucleon system \cite{Hammer:2019poc},
when the subleading interactions are treated in distorted-wave perturbation
theory to ensure renormalization 
\cite{vanKolck:1997ut,Kaplan:1998tg,Kaplan:1998we,vanKolck:1998bw}. 
Having incorporated the unnaturalness at LO, we might expect 
that corrections will contain no additional failure
of naturalness, so that subleading LECs can be estimated from loop
diagrams by $\Lambda \to M_{\rm hi}$.
For example, the $\Lambda^{-1}$ term in Eq. \eqref{EREparamsEFT} leads to 
$C_{2}$ at NLO as given by Eq. \eqref{fineDANN}
--- this is one order down the expansion, in contrast to (modified) NDA.
Equivalently, $|r_2|={\cal O}(M_{\rm hi}^{-1})$ just as expected from naturalness.
The argument continues at higher order.

For $|k|\ll |\kappa_2|$
we can Taylor-expand the denominator of Eq. \eqref{Tsum} to obtain 
Eq. \eqref{TNNpert}. As far as the first two terms of the latter are
concerned, it seems as if we are back to the naturalness case with 
$M_{\rm hi}=\kappa_2$, 
but effective-range parameters other than the scattering
length would look small.
A better quantum-field theoretical example to replace
a nonrelativistic potential as a toy underlying theory
is a Pionless EFT for two coupled channels \cite{Cohen:2004kf,Braaten:2007nq},
which models a Feshbach resonance.
When all entries of the $2\times 2$ matrix that replaces $C_{0}$
have about the same magnitude $\kappa_R^{-1}$ and the 
channel thresholds differ in energy by $\Delta E \sim \kappa_R^2/2\mu$ 
(where $\mu$ is the
reduced mass), one generally 
finds $|a_2|\sim |r_2|\sim \kappa_R^{-1}$ in the open channel
--- the natural situation.
By dialing $\kappa_R^{-1}$ against a combination of entries of the 
$2\times 2$ coupling matrix, one can produce $|a_2|\gg  \kappa_R^{-1}$
while $r_2\sim -\kappa_R^{-1}$ --- the fine-tuned scenario.

Now that we have an EFT that accounts for
shallow two-nucleon states, 
we may ask what the consequences 
for larger nuclei are. The answer is, lots of surprises.
The first surprise was probably the observation by
Thomas \cite{Thomas:1935zz} of a ``collapse'' of the $A=3$ system. 
With the LO two-body interactions, 
the ground-state binding energy is $B_{3}\propto \Lambda^2/m_N$ for 
$\Lambda\gg |\kappa_2|$ 
and, as $\Lambda$ increases, 
excited bound states emerge and collapse as well.
The half-off-shell
amplitude for scattering of a particle on the two-body bound state
displays a bizarre behavior:
it oscillates as a function of the off-shell momentum with
a phase that depends on $\ln \Lambda$ 
\cite{Bedaque:1998kg,Bedaque:1998km,Bedaque:1999ve}.
Small cutoff variations result in large changes at low momentum,
a regulator dependence that indicates that the 
$A=3$ system is not renormalizable with only $C_{0}$ interactions.
The same conclusion holds for bosons, which also suffer from the absence
of the Pauli exclusion principle. 

The second surprise is that,
since two-body interactions with more derivatives are small
within the range of the EFT, the appropriate counterterm must be a 
three-body force. And an unexpected one at that.
Indeed, the $D_0$ term in Eq. \eqref{pilessLag}
can exactly counterbalance
the cutoff variation 
if \cite{Bedaque:1998kg,Bedaque:1998km,Bedaque:1999ve}
\begin{equation}
D_0(\Lambda)\propto \frac{1}{\Lambda^4}
\frac{\sin \!\left(s_0 \ln(\Lambda/\Lambda_\star) -\arctan s_0^{-1}\right)}
{\sin \!\left(s_0 \ln(\Lambda/\Lambda_\star) +\arctan s_0^{-1}\right)}
\left(1+ {\cal O}\left(\kappa_2/\Lambda\right)\right),
\label{D0}
\end{equation}
where $s_0 \simeq 1.00624$ and $\Lambda_\star$ is a physical parameter.
Once one low-momentum datum is reproduced by a choice of $\Lambda_\star$,
the phase of the half-off-shell scattering amplitude 
is fixed and all other low-momentum observables attain finite values
as $\Lambda$ increases.
This is true, in particular, of bound states. 
Instead of the periodic emergence of bound states at zero energy
before renormalization, after renormalization one
observes the periodic emergence of {\it deeper} bound states,
which achieve finite binding energies as $\Lambda$ increases.
The bound states that appear once $\Lambda \simge M_{\rm hi}$ 
are unphysical because they are beyond the range of validity of the EFT.

Unlike the two-derivative two-body force, the two-derivative three-body force
with LEC $D_2$ (to be found in the ``$\dots$'' of Eq. \eqref{pilessLag})
appears at N$^2$LO
\cite{Bedaque:1998km,Hammer:2001gh,Bedaque:2002yg,Platter:2008cx,Ji:2012nj}.
Since $D_0$ is an LO interaction, the three-body LECs effectively scale as 
\begin{equation}
|D_0|={\cal O}\left(M_{\rm lo}^{-4}\right), 
\quad
|D_2|={\cal O}\left(M_{\rm lo}^{-4}M_{\rm hi}^{-2}\right).
\label{fineDA3N}
\end{equation} 
This is very different from what is expected from modified NDA
with natural-sized $C_{0}$.
Consider the two-loop three-body diagram with four successive $C_{0}$ 
interactions involving each time a different pair of particles.
Under the rules \eqref{red2} and \eqref{red1}, 
it generates a $\ln \Lambda$ dependence, whose coefficient leads to
the scaling of $D_0$ in Eq. \eqref{modNDANN}.
Conversely, an unnatural $C_{0}$ induces an unnatural $D_0$. 
However, while this diagram for unnatural $C_{0}$ 
suggests that $M_{\rm lo}$ is determined by $\kappa_2$, 
the nonperturbative renormalization that leads to Eq. \eqref{D0} 
means the three-body energies are fixed by $\Lambda_\star$.
$M_{\rm lo}$ is set by $\kappa_3$.

There is no relative renormalization enhancement of higher-body forces at LO,
as the calculations of
Refs. \cite{Platter:2004he,Platter:2004zs,Stetcu:2006ey,Hammer:2006ct,Kirscher:2009aj,Kirscher:2015yda,Bazak:2016wxm,Contessi:2017rww} found convergence in
$A=4,5,6$ binding energies as $\Lambda$ increases.
An enhancement takes place at NLO, though, so that a new scale appears
through a four-body force \cite{Bazak:2018qnu}
\begin{equation}
|E_0|={\cal O}\left(M_{\rm lo}^{-6} M_{\rm hi}^{-1}\right).
\label{fineDA4N}
\end{equation} 
Once this force
is accounted for, no more-body forces are needed for NLO renormalization
in larger systems.

Therefore, at LO there is a single parameter $\Lambda_\star$ 
not determined by $A=2$ physics.
Not every $A\ge 3$ observable is sensitive to this parameter.
For example, neutron-deuteron ($nd$)
scattering in the spin-3/2 channel, where the two neutron spins are aligned,
can be predicted to a very good accuracy from $A=2$ physics 
\cite{Bedaque:1997qi,Bedaque:1998mb,Vanasse:2013sda}.
But correlations should exist through $\Lambda_\star$ 
among $A\ge 3$ observables not affected by the exclusion principle.
The classic example is the Phillips line \cite{Phillips:1968zze}
on the plane spanned by the triton binding energy
and the spin-1/2 $nd$ scattering length.
This correlation was first discovered empirically
with points representing various phenomenological 
potentials, which describe two-nucleon data up to relatively high momenta.
In the EFT, this correlation is produced 
as $\Lambda_\star$ is varied
\cite{Bedaque:1998kg,Bedaque:1998km,Bedaque:1999ve}.
The EFT line lies close not only to the experimental point
but also to the empirical line,
meaning that the  many parameters of the various phenomenological potentials
amount to a single relevant parameter $\Lambda_\star$.
Similarly, $A\ge 4$ ground-state binding energies are correlated
with the $A=3$ ground-state energy 
\cite{Platter:2004he,Platter:2004zs,Bazak:2016wxm}
in a Tjon \cite{Tjon:1975sme,NakAkaTanLim78} and 
generalized Tjon \cite{NakLimAkaTan79,LimNakAkaTan80} lines.

$\slashpi$EFT bags many other successes for 
light nuclei \cite{Hammer:2019poc}.
It is not currently known how far in $A$ $\slashpi$EFT can accommodate
the growing nuclear binding energies. 
But within its regime of validity we should 
be able to derive nuclear properties with the contact
interactions in Eq. \eqref{pilessLag}.  
The small scales they contain will contaminate heavier nuclei and 
lead to anomalously small energies or energy gaps. 
The application of $\slashpi$EFT to $A\ge 5$ nuclei is in its infancy.
The first indication \cite{Stetcu:2006ey,Contessi:2017rww,Bansal:2017pwn} 
is that at LO a gas of nucleons and alpha particles results. 
Whether subleading corrections will lead to the relatively small
binding (relative to these nucleon-alpha thresholds) is an open question.
We might conjecture that,
in addition to purely accidental near coincidences among the large number of 
nuclear excited states and thresholds, unnatural scales
in heavier nuclei trace back to those in light nuclei.

\section{Unraveling unnaturalness}
\label{symmetry}

Is nuclear physics {\it technically} unnatural, though? 
In this section I discuss some of 
the evidence that suggests the answer might be ``no''.
I will argue that there is an approximate symmetry that ensures that 
a small scale enters nuclear physics.

In the scale of the triton binding binding momentum
$\kappa_3\simeq 70$ MeV, the $^1S_0$ two-nucleon
pole is very shallow, and even the deuteron binding momentum is 
somewhat small.
This suggests that the unitarity limit $\kappa_2\to 0$ 
in Eq. \eqref{Tsum} might be a good approximation for the physics
of the ground states of larger nuclei 
\cite{Konig:2015aka,Konig:2016utl,Konig:2016iny,Kolck:2017zzf,Konig:2019xxk}.
(See also Refs. \cite{Kievsky:2018xsl,Gattobigio:2019omi}
for a related approach.)
In this limit, which is a non-trivial fixed point 
of the renormalization group (RG) \cite{Weinberg:1991um},
the LO two-nucleon amplitude \eqref{Tsum}
has only the unitarity term $ik$, with no
dimensionful parameter.

The vanishing of two-body binding energies in the unitarity limit 
is a reflection of scale invariance. 
Under a change of scales \cite{Hagen:1972pd}
with parameter
$\alpha>0$,
\begin{equation}
r\to \alpha r, 
\qquad
t/m_\psi\to \alpha^2 t/m_\psi, 
\qquad
\Lambda\to \alpha^{-1} \Lambda,
\qquad
\psi \to \alpha^{-3/2} \psi,
\label{scale}
\end{equation}
the nucleon kinetic and $C_{0}$
terms in Eq. \eqref{pilessLag} are invariant, but only when $\kappa_2\to 0$ 
in Eq. \eqref{C00}. 
Under a scale change, $mE\to \alpha^{-2}mE$ but in the unitarity limit
there is no two-body scale, so $B_2$ must vanish. 
In this limit the $A=2$ system is also conformally invariant 
\cite{Mehen:1999nd}.

Beyond the two-body system,
scale (as well as conformal) invariance is ``anomalously'' broken by
the dimensionful parameter $\Lambda_\star$.
The latter arises from renormalization, 
even though at unitarity we start without any scale,
so we could call it dimensional transmutation.
The dependence of $D_0$ on $\Lambda$ in Eq. \eqref{D0} reveals an
RG limit cycle
\cite{Bedaque:1998kg,Bedaque:1998km,Bedaque:1999ve}.
As a consequence, all LO terms
in Eq. \eqref{pilessLag} are invariant under a transformation
\eqref{scale}, but only for discrete values
\begin{equation}
\alpha_l = e^{l\pi/s_0}\simeq (22.7)^{l},
\label{disscale}
\end{equation}
with $l$ an integer.
The limit cycle engenders discrete scale invariance (DSI) 
\cite{Sornette:1997pb} which is present for 
all $A$ within the range of $\slashpi$EFT.

Since it is the only dimensionful parameter at two-body unitarity,
$\Lambda_\star$ sets the scale for {\it all} $A\ge 3$ binding energies. 
For ground states, eliminating $\Lambda_\star$ translates into a universal 
form \cite{Carlson:2017txq}
for the correlations mentioned in the previous section,
\begin{equation}
\frac{B_A}{A} = \xi_A \frac{B_3}{3}.
\label{universal}
\end{equation}
The numbers $\xi_A$ are universal in the sense that they are the same
for any type of unitary four-component fermion. 
The same relation holds for bosons with the same value for $\xi_4$
but different $\xi_{A\ge 5}$ on account of the absence of
the exclusion principle.
The value $\xi_4\simeq 3.46$ is well established \cite{Deltuva:2010xd}.
For bosons we find at small $A$ \cite{Bazak:2016wxm}
\begin{equation}
\xi_A \approx 3A \left(1-\frac{2}{A}\right)^2,
\label{universalfewbos}
\end{equation}
while for large $A$ \cite{Carlson:2017txq}
\begin{equation}
\xi_A \approx \xi_\infty\left(1-\frac{\eta}{A^{1/3}}+\ldots \right),
\label{universalmanybos}
\end{equation}
with $\xi_\infty\simeq 90 \pm 10$ and $\eta\simeq 1.7\pm 0.3$.
Many bosons at unitarity thus approximately satisfy, like nuclei,
the liquid-drop formula and can be thought of
forming a quantum liquid. 
The behavior is similar to that of $^4$He atomic clusters 
\cite{Pandharipande:1983}.

For nucleons, not much is known. When corrections linear in $a_2^{-1}$ 
are included, 
$\xi_4$ becomes $\simeq 2.6$ \cite{Konig:2016utl}.
This is very close to the experimental
value $\xi_4^{\rm exp}\simeq 2.5$ obtained from the triton and alpha
binding energies. A full NLO calculation involves the four-body force
with LEC $E_0$ \cite{Bazak:2018qnu}, which can be adjusted to reproduce
$\xi_4^{\rm exp}$.
A small number ($A=8$)
of unitary four-component fermions tends to cluster into alpha-like
objects \cite{Dawkins:2019vcr}, as observed in $^8$Be. 
The behavior for other $A$ is virgin territory, even at LO.

DSI has strong consequences also for excited states. 
For $A=3$ I write binding energies as
\begin{equation}
B_{3;n} = \beta_{3;n} \, B_3,
\label{B3n}
\end{equation}
with $n\ge 0$ and $\beta_{3;0}=1$.
The $mE\to \alpha_l^{-2}mE$ transformation under a scale change
can now support non-vanishing energies if
\begin{equation}
\beta_{3;n}= e^{-2n\pi/s_0}.
\label{beta3nunit}
\end{equation}
This is Efimov's famous geometric tower 
\cite{Efimov:1970zz,Efimov:1971zz},
which extends up to threshold.
An extensive review of the properties of these states can be found
in Ref. \cite{Braaten:2004rn}.
As two-body attraction decreases and $\kappa_2$ becomes negative, 
one observes the amazing quantum phenomenon
of Borromean binding, where the three-body system shows bound states while
no two-body bound state exists.
If, instead, attraction increases and $\kappa_2$ 
becomes positive, three-body bound states {\it dis}appear into
the particle-dimer threshold.
In either case, only a few of these states are physical.
For atomic $^4$He, both ground \cite{Schoellkopf:1996} and 
first-excited \cite{Kunitski:2015qth} states have been detected.
For nucleons, 
only the ground state (triton/helion) is observed, 
but there is a virtual state in $nd$ scattering 
that becomes the triton excited state as $\kappa_{20}$ is 
decreased \cite{Rupak:2018gnc}.

For $A=4$, there are two states for each Efimov state 
\cite{Hammer:2006ct},
one $1.002$ more bound than the $A=3$ parent, the other 
$\simeq 4.6$ times deeper \cite{Deltuva:2010xd}. 
Remarkably, potential-model calculations 
\cite{vonStecher:2009qw,Gattobigio:2011ey,vonStecher:2011zz,Gattobigio:2012tk}
show that
this doubling process continues with increasing number of {\it bosons}.
For a given $A$, there are $2^{(A-3)}$ interlocking towers of states.
The replicating towers are a reflection of DSI,
but the doubling, which has a topological interpretation 
\cite{Horinouchi:2016gqg}, is of unclear origin.
For $A\ge 5$ four-component fermions the pattern of towers is not known.

The lowest $A=4$ Efimov-state descendants have been spotted in atomic systems 
\cite{Ferlaino:2009zz}.
The first excited state of the alpha particle is close to the
nucleon-triton threshold, 
another evidence that light nuclei are perturbatively 
close to unitarity \cite{Konig:2016utl}.
This state can be thought of as a two-body, nucleon-triton ``halo'' state,
since its separation energy is much smaller than the triton binding energy.
One would have thought this situation could only arise from fine tuning!
Yet, each top boson state in the doubling is automatically a halo state
consisting of a boson and an $A-1$ boson cluster.
It is possible that similar structures arise for five or more unitary 
four-component fermions, and indeed they are observed in ``halo nuclei''.
The poster child is $^6$He, which is 
only $\simeq 1.04$ times more bound than the alpha particle
and can be viewed as a three-body state of two neutrons and one alpha.
It could thus be that even such an apparent fine tuning has
an origin in DSI.
Regardless of the cause, halo states can be described by an EFT,
Halo EFT \cite{Bertulani:2002sz,Bedaque:2003wa}, in a way that
is completely parallel to $\slashpi$EFT for nucleons,
only with additional fields for tight clusters of nucleons
such as the alpha particle.

\section{Unclear naturalness?}
\label{conc}

I have emphasized that some notion of naturalness is necessary
for power counting, which in turn is essential for EFT predictive power.
In perturbative theories, naive dimensional analysis provides 
a reasonable definition of naturalness.
I went on to show that a modification of NDA needed for heavy fermion
systems implies that light nuclei are contaminated by a small scale.
I linked this small scale to approximate discrete scale invariance,
which, to the extent it holds, ensures the properties of heavier
nuclei are tied to those of the three-body system.
I even speculated, based on calculations for bosons, that
some coincidences observed in nuclear physics could, perhaps,
be generated automatically.

Still, there are some clouds on the horizon.   
First, due to the nonperturbative nature of nuclei,
certain interactions in the effective Lagrangian \eqref{pilessLag}
are enhanced in the limit DSI is exact, rather than suppressed 
as per 't Hooft's expectation. This is not necessarily a problem,
as the introduction of auxiliary fields shows.
To handle the shallow states we can introduce
two ``dimer'' fields $D_I$ \cite{Kaplan:1996nv} 
and one ``trimer'' field $T$ \cite{Bedaque:2002yg}.
Still leaving the two channels $I=0,1$ implicit,
\begin{eqnarray}
{\cal L}_{\minislashpi\rm{EFT}} &=& 
N^\dagger \left(i \partial_0 + \frac{\vec \nabla^2}{2m_N} \right) N
\nonumber\\
&& 
+ {D}^\dagger \left[-\Delta
+ \delta \left(i m_N \partial_0 + \frac{\vec \nabla^2}{4}\right) \right] D
-\sqrt{\frac{4\pi}{m_N}}
\left(N^\dagger N^\dagger D + \mathrm{H.c.}\right)
\nonumber\\
&& 
+ {T}^\dagger \left[-\Omega
+ \omega \left(i m_N \partial_0 + \frac{\vec \nabla^2}{6}+\frac{\kappa_2^2}{m_N}
\right)\right] T
- \sqrt{4\pi}\left(N^\dagger D^\dagger \, T+ \mathrm{H.c.}\right)
\nonumber\\
&& + 4\pi \, \Theta_0 \, T^\dagger T \, N^\dagger N
+ \ldots,
 \label{auxLag}
\end{eqnarray}
where 
$\Delta=C_{0}^{-1}$, $\delta=C_{0}^{-2}C_{2}$, 
$\Omega=C_{0}^{2}D_{0}^{-1}$, $\omega=C_{0}^{2}D_{0}^{-2}D_{2}$, 
$\Theta_{0}= C_{0}^{2}D_{0}^{-2} E_0$, {\it etc.}
As the only two-body parameter with non-negative mass dimension ($+1$), 
one might expect the scale-invariance suppression 
$\Delta^{\rm red}={\cal O}(\kappa_2/M_{\rm hi})$ in the dimeron residual mass.
It resembles the effect of symmetries on other masses
--- {\it e.g.} on the nucleon mass difference \eqref{chEFTNDA}
that appears in the Lagrangian \eqref{chiLag} --- and it
is indeed sufficient to convert
the naturalness relation \eqref{modNDANN} into the near-unitarity 
expectation \eqref{fineDANN}.
For the three-body parameters with non-negative mass dimensions 
($+2$ and $0$), we have to assume
$\Omega^{\rm red}={\cal O}(M_{\rm lo}^2/M_{\rm hi}^2)=\omega^{\rm red}$,
which then yields the relations \eqref{fineDA3N} and \eqref{fineDA4N}.
The ratio $M_{\rm lo}/M_{\rm hi}\ll 1$ must somehow arise once 
$\kappa_2/M_{\rm hi}\ll 1$.

Second, it is unclear how the symmetry emerges in QCD or even $\chi$EFT.
The most important element of $\chi$EFT in nuclear physics is 
one-pion exchange (OPE).
At $Q\sim m_\pi$, OPE has a magnitude 
\cite{Kaplan:1998tg,Kaplan:1998we}
\begin{equation}
|V_{\rm OPE}|= {\cal O}\left(4\pi m_N^{-1} M_{N\!N}^{-1}\right),
\quad
M_{N\!N} \equiv \frac{16\pi f_\pi^2}{g_A^2 m_N}.
\label{OPEscale}
\end{equation}
While NDA implies $M_{N\!N}={\cal O}(f_\pi)$, numerically $M_{N\!N}\simeq 290$ MeV.
For $Q\sim M_{N\!N}$, pions are nonperturbative.
OPE is a singular potential in spin-triplet waves, and its
nonperturbative renormalization requires a 
chirally symmetric LEC in each wave 
where OPE is attractive \cite{Nogga:2005hy}.
The solution of the Schr\"odinger equation oscillates with
a wavelength determined by $M_{N\!N}$ while the
LEC fixes the asymptotic phase \cite{Beane:2000wh}.
Although OPE is not singular in the $^1S_0$ channel, its interference with the 
chirally symmetric contact interaction demands an 
additional chiral-symmetry-breaking LEC for 
renormalization \cite{Kaplan:1996xu}.
In both $S$-wave channels it is thus a combination of OPE
and LEC that must enforce shallow two-nucleon poles.
But $M_{N\!N}$ remains and scale invariance is apparently broken explicitly.
Scale invariance must be an emergent symmetry 
for $Q\simle M_{N\!N}$, where pion exchange should be perturbative
\cite{Kaplan:1998tg,Kaplan:1998we}.
In this case, DSI manifests itself just as in $\slashpi$EFT.
Unfortunately the perturbative-pion expansion breaks
down already below $M_{N\!N}$, at least in the $^3S_1$ and $^3P_0$ channels
\cite{Fleming:1999ee,Kaplan:2019znu}.
The convergence for $^1S_0$ is also slow at best \cite{Beane:2001bc}.
The numerical character of nonperturbative OPE calculations
obscures the emergence of DSI.

There is much to do, both in exploring the manifestations of DSI 
in larger nuclei and in understanding its emergence in QCD.
Still, despite the current challenges in particle physics,
a criterion of naturalness supplies an essential ingredient for nuclear EFTs.

\section*{Acknowledgments}
I thank Matt Baumgart, Ozan Erdogan, and Jaber Balal Habashi 
for useful discussions.
This material is based upon work supported in part 
by the U.S. Department of Energy, Office of Science, Office of Nuclear Physics, 
under award DE-FG02-04ER41338
and by the European Union Research and Innovation program Horizon 2020
under grant No. 654002.

\end{document}